# Analysis of Block OMP using Block RIP

**Jun Wang, Gang Li, Hao Zhang, Xiqin Wang**

Department of Electronic Engineering, Tsinghua University, Beijing 100084, China

Emails: jun-wang05@mails.tsinghua.edu.cn, {gangli, haozhang, wangxq_ee}@tsinghua.edu.cn

## Abstract

Orthogonal matching pursuit (OMP) is a canonical greedy algorithm for sparse signal reconstruction. When the signal of interest is block sparse, i.e., it has nonzero coefficients occurring in clusters, the block version of OMP algorithm (i.e., Block OMP) outperforms the conventional OMP. In this paper, we demonstrate that a new notion of block restricted isometry property (Block RIP), which is less stringent than standard restricted isometry property (RIP), can be used for a very straightforward analysis of Block OMP. It is demonstrated that Block OMP can exactly recover any block $K$-sparse signal in no more than $K$ steps if the Block RIP of order $K+1$ with a sufficiently small isometry constant is satisfied. Using this result it can be proved that Block OMP can yield better reconstruction properties than the conventional OMP when the signal is block sparse.

## I. Introduction

Recently, the compressed sensing (CS) theory [1, 2] has been used in many areas because of its excellent performance on reconstruction of sparse signals. Consider an equation $\mathbf{y} = \mathbf{D}\mathbf{x}$, where $\mathbf{x}$ is a $K$-sparse signal of length $N$ to be determined, $\mathbf{D}$ is a $L \times N$ measurement matrix (where $L < N$ typically), $\mathbf{y}$ is the measurement vector of length $L$. Many algorithms have been proposed to determine $\mathbf{x}$ in a stable and efficient manner. Orthogonal matching pursuit (OMP) is one of conventional greedy algorithms, which is commonly used in CS area for its simplicity. Two theoretical tools have been used to analyze OMP algorithm [3, 4]. One is noted as coherence

[3]: $\mu \triangleq \max_{i \neq j} \frac{|\langle \mathbf{d}_i, \mathbf{d}_j \rangle|}{\|\mathbf{d}_i\|_2 \|\mathbf{d}_j\|_2}$, where $\mathbf{d}_i$ denotes the *i*th column of the matrix $\mathbf{D}$. When the coherence of measurement matrix satisfies that $u < \frac{1}{2K-1}$, it has been shown that OMP can recover any *K*-sparse signal from the measurements $\mathbf{y}$ [3]. The other one is restricted isometry property (RIP), it has been shown that OMP can exactly recover any *K*-sparse signal in no more than *K* steps if the RIP of order *K*+1 with isometry constant $\delta < \frac{1}{3\sqrt{K}}$ is satisfied [4].

In this paper we consider block sparse signals that have nonzero coefficients occurring in clusters. The problem of interest is that whether explicitly taking the block sparsity into account can yield better reconstruction properties than treating the signal as a conventional sparse signal. This problem is treated in [5], where it is shown that, a mixed $l_2/l_1$-norm algorithm for recovering block-sparse signals can yield exact sparse recovery if the block restricted isometry property (Block RIP) [5] is satisfied. In [6], a notion of the block coherence is proposed, and the exact sparse recovery can be achieved by using the Block OMP algorithm when the block coherence is sufficiently small. For the more general setting of model-based compressed sensing where block-sparsity is included as a special case, it is shown in [7] that a modified vision of CoSaMP algorithm [8] can yield excellent recovery reconstruction properties for block sparse signals.

In this paper we analyze the recovery performance of Block OMP using Block RIP, which has not been studied in existing literature. Our approach is partly based on [4] and [9] (and the mathematical techniques used therein). The main contribution in this paper is proving that, if the dictionary matrix $\mathbf{D}$ satisfies the Block-RIP of order *K*+1 with a sufficiently small isometry constant, the Block OMP can exactly recover any block *K*-sparse signal in no more than *K* steps, which is looser than the exact recovery condition using the conventional OMP. Moreover, it can also be indicated that the iteration steps using Block OMP are less than using OMP. This paper is organized as follows. Foundational concepts of block sparse signal reconstruction , i.e., block sparsity, Block RIP and Block OMP are reviewed in Section II. The analysis of Block OMP using Block RIP is performed in Section III. Finally, a brief conclusion is given in Section IV.

*Notation:* Vectors and matrices are denoted by boldface letters. All vectors are column vectors. $(\cdot)^T$ denotes the transpose operation; $\|\cdot\|_1$ and $\|\cdot\|_2$ denotes $l_1$ and $l_2$ norms, respectively.

## II. Review of block sparse signal reconstruction

In this section, we review some foundational concepts of block sparse signal reconstruction, including block sparsity, Block RIP and Block OMP.

### A. Block Sparsity

Assume that $\mathbf{x}$ consists of blocks with given block length $d$, and $N = Md$ with an integer $M$, then $\mathbf{x}$ can be expressed as:

$$\mathbf{x} = [\underbrace{x_1; \cdots, x_d}_{\mathbf{x}^T[1]}, \underbrace{x_{d+1}; \cdots, x_{2d}}_{\mathbf{x}^T[2]}, \cdots, \underbrace{x_{N-d+1}; \cdots, x_N}_{\mathbf{x}^T[M]}], \tag{1}$$

where $\mathbf{x}[l]$ denotes a vector which starts from the $(l-1)d+1$ th element and ends to the $ld$ th element of $\mathbf{x}$, $d$ is an integer, $\mathbf{x}[l]$ is called the $l$th block of $\mathbf{x}$. The signal $\mathbf{x}$ is called a $d$-block $K$-sparse, if the number of $\mathbf{x}[l]$ which has nonzero $l_2$ norm is no more than $K$. Using the definition of mixed $l_2/l_0$-norm [6], block sparsity can also be expressed as:

$$\|\mathbf{x}\|_{2,0} = \sum_{l=1}^{M} I(\|\mathbf{x}[l]\|_2 > 0), \tag{2}$$

where $I(\|\mathbf{x}[l]\|_2 > 0)$ is a indicator function, which is 1 if $\|\mathbf{x}[l]\|_2 > 0$ and 0 otherwise.

### B. Block RIP

First we review the conventional RIP [1, 2]. A matrix $\mathbf{D} \in \mathbb{R}^{L \times N}$ satisfies the RIP of order $K$ if there is a constant $\delta \in (0,1)$ such that:

$$(1-\delta)\|\mathbf{x}\|_2^2 \leq \|\mathbf{D}\mathbf{x}\|_2^2 \leq (1+\delta)\|\mathbf{x}\|_2^2, \tag{3}$$

for all $K$-sparse $\mathbf{x} \in \mathbb{R}^N$. Ref. [5] extends this property to block-sparse vectors and lead to the following definition. The matrix $\mathbf{D}$ has the Block RIP of order $K$ with isometry constant $\delta_d \in (0,1)$, if for all $d$-block $K$-sparse $\mathbf{x} \in \mathbb{R}^N$, we have

$$(1-\delta_d)\|\mathbf{x}\|_2^2 \leq \|\mathbf{D}\mathbf{x}\|_2^2 \leq (1+\delta_d)\|\mathbf{x}\|_2^2. \tag{4}$$

Note that a $d$-block $K$-sparse vector is $Kd$-sparse in the conventional sense. If $\mathbf{D}$ satisfies the RIP of order $Kd$

with isometry constant $\delta$, (3) must hold for all $d$-block $K$-sparse $\mathbf{x} \in \mathbb{R}^N$. On the contrary, if $\mathbf{D}$ satisfies the Block RIP of order $K$ with isometry constant $\delta_d$, (4) may not hold for all $Kd$-sparse $\mathbf{x} \in \mathbb{R}^N$. As a result, the RIP of order $Kd$ is the sufficient condition of Block RIP of order $K$ for the same constant $\delta$. In the other word, the Block RIP constant $\delta_d$ is typically smaller than the conventional RIP constant $\delta$ for the same matrix.

### C. Block OMP algorithm

Block OMP is the block vision of OMP, and accomplished in reconstruct block sparse signals [6]. The entire algorithm is specified in Algorithm 1.

---

**Algorithm 1 Block Orthogonal Matching Pursuit**

---

**input:** measurement matrix $\mathbf{D}$, measurements $\mathbf{y}$, stopping iteration index $L$

**initialize:** residual error $\mathbf{r}^0 = \mathbf{y}$, signal $\mathbf{x}^0 = \mathbf{0}$, support set $\Lambda^0 = \varnothing$, iteration index $l = 0$

**while** $l < L$

1. $\mathbf{h}^l = \mathbf{D}^T \mathbf{r}^l$

2. $\Lambda^{l+1} = \Lambda^l \cup \{\arg\max_j \|\mathbf{h}^l[j]\|_2\}$

3. $\mathbf{x}^{l+1} = \arg\min_{\mathbf{z}:\text{supp}_B(\mathbf{z}) \subseteq \Lambda^{l+1}} \|\mathbf{y} - \mathbf{Dz}\|_2$

$\mathbf{r}^{l+1} = \mathbf{y} - \mathbf{D}\mathbf{x}^{l+1}$

$l = l + 1$

**end**

**output:** $\hat{\mathbf{x}} = \mathbf{x}^L$

where $\text{supp}_B(\mathbf{z}) \triangleq \{l \in \{1, 2, \cdots, M\} \mid \|\mathbf{z}[l]\|_2 \neq 0\}$, and $\Lambda^l, \mathbf{h}^l, \mathbf{r}^l, \mathbf{x}^l$ denotes $\Lambda, \mathbf{h}, \mathbf{r}, \mathbf{x}$ at the $l$th iteration, respectively.

Compared with the conventional OMP algorithm, the main difference of Block OMP is that Block OMP chooses the block index according to $\arg\max_j \|\mathbf{h}^l[j]\|_2$, while the conventional OMP chooses the index according to $\arg\max_j |\mathbf{h}^l_j|$, where $\mathbf{h}^l[j]$ is a vector and $\mathbf{h}^l_j$ is a scalar.

## III. Analysis of Block OMP using Block RIP

In this section, we begin with some observations regarding Block OMP which set the stage for our main results. These results include four lemmas, two corollaries and a theorem. Lemma 1 and Lemma 2 is the block vision of Lemma 3.1 and Lemma 3.2 in [4], Then we prove Lemma 3, Lemma 4, Corollary 1 and Corollary 2, and at last prove Theorem 1, which induces the main conclusion of this paper.

Similarly to analysis of OMP in [4], our goal is to find how to guarantee that the block index chosen at each iteration step is correct. Let $\Re(\mathbf{D}_\Lambda)$ denote the spanning space of columns of $\mathbf{D}_\Lambda$, where $\mathbf{D}_\Lambda$ is a $L \times |\Lambda|d$ matrix and is defined as $\mathbf{D}_\Lambda \triangleq [\mathbf{D}[\Lambda(1)], \mathbf{D}[\Lambda(2)], \cdots, \mathbf{D}[\Lambda(|\Lambda|)]]$ (here $|\Lambda|$ is the length of $\Lambda$). Before the iteration stops, it is accepted that $\mathbf{D}_\Lambda$ is a matrix of full column rank ($|\Lambda|d \leq L$), then the Moore-Penrose pseudoinverse of $\mathbf{D}_\Lambda$ can be denoted by $\mathbf{D}_\Lambda^\dagger = (\mathbf{D}_\Lambda^T \mathbf{D}_\Lambda)^{-1} \mathbf{D}_\Lambda^T$. It is clear to see that the orthogonal projection operator onto $\Re(\mathbf{D}_\Lambda)$ is $\mathbf{P}_\Lambda = \mathbf{D}_\Lambda \mathbf{D}_\Lambda^\dagger$, while the orthogonal projection operator onto the orthogonal complement of $\Re(\mathbf{D}_\Lambda)$ is $\mathbf{P}_\Lambda^\perp = (\mathbf{I} - \mathbf{P}_\Lambda)$. Then we define $\mathbf{A}_\Lambda$ as $\mathbf{A}_\Lambda \triangleq \mathbf{P}_\Lambda^\perp \mathbf{D}$, which is the result of orthogonalizing the columns of $\mathbf{D}$ against $\Re(\mathbf{D}_\Lambda)$. It is easy to prove that blocks of $\mathbf{A}_\Lambda$ indexed by $\Lambda$ are equal to zeros.

Now we consider how $\mathbf{x}^i$ is obtained at the *i*th iteration in Algorithm 1. $\mathbf{x}^i$ is solved as a least squares problem, so it is given by:

$$\mathbf{x}^i|_{\Lambda^i} = \mathbf{D}_{\Lambda^i}^\dagger \mathbf{y}, \text{ and } \mathbf{x}^i|_{(\Lambda^i)^c} = 0, \tag{5}$$

where $\mathbf{x}|_\Lambda \in \mathbb{R}^{|\Lambda|d}$ is defined as $\mathbf{x}|_\Lambda \triangleq [\mathbf{x}^T[\Lambda(1)], \mathbf{x}^T[\Lambda(2)], \cdots, \mathbf{x}^T[\Lambda(|\Lambda|)]]^T$, and $\Lambda^c \triangleq \{1, 2, \cdots, M\} / \Lambda$.

Since $\mathbf{P}_{\Lambda^i}^\perp$ is orthogonal projection operator, $\mathbf{P}_{\Lambda^i}^\perp = (\mathbf{P}_{\Lambda^i}^\perp)^T = (\mathbf{P}_{\Lambda^i}^\perp)^2$. Then we have

$$\begin{aligned} \mathbf{r}^i &= \mathbf{y} - \mathbf{D}\mathbf{x}^i = \mathbf{y} - \mathbf{D}_{\Lambda^i} \mathbf{D}_{\Lambda^i}^\dagger \mathbf{y} \\ &= (\mathbf{I} - \mathbf{P}_{\Lambda^i})\mathbf{y} = \mathbf{P}_{\Lambda^i}^\perp \mathbf{y} = \mathbf{P}_{\Lambda^i}^\perp \mathbf{D}\mathbf{x} = \mathbf{A}_{\Lambda^i} \mathbf{x} \end{aligned}, \tag{6}$$

and

$$\mathbf{r}^i = \mathbf{P}_{\Lambda^i}^\perp \mathbf{y} = \mathbf{P}_{\Lambda^i}^\perp \mathbf{P}_{\Lambda^i}^\perp \mathbf{y} = (\mathbf{P}_{\Lambda^i}^\perp)^T \mathbf{P}_{\Lambda^i}^\perp \mathbf{y}, \tag{7}$$

so

$$\mathbf{h}^i = \mathbf{D}^T \mathbf{r}^i = \mathbf{D}^T (\mathbf{P}_{\Lambda^i}^\perp)^T \mathbf{P}_{\Lambda^i}^\perp \mathbf{y} = \mathbf{A}_{\Lambda^i}^T \mathbf{r}^i = \mathbf{A}_{\Lambda^i}^T \mathbf{A}_{\Lambda^i} \mathbf{x}. \tag{8}$$

We expect that $\mathbf{h}^i$ can hold most of information of $\mathbf{x}$, which means that $\|\mathbf{h}^i - \mathbf{x}\|_2$ is expected not too large. Lemma 3 below will prove the correctness of this hypothesis.

We begin with two lemmas which are just the block version of Lemma 3.1 and Lemma 3.2 in [4].

**Lemma 1** [4]: Suppose $\mathbf{u}, \mathbf{v} \in \mathbb{R}^N$, if $\mathbf{D}$ has the Block RIP of order $K = \max(\|\mathbf{u}+\mathbf{v}\|_{2,0}, \|\mathbf{u}-\mathbf{v}\|_{2,0})$ (see (2)) with isometry constant $\delta_d \in (0,1)$, then:

$$|\langle \mathbf{Du}, \mathbf{Dv}\rangle - \langle \mathbf{u}, \mathbf{v}\rangle| \leq \delta_d \|\mathbf{u}\|_2 \|\mathbf{v}\|_2. \tag{9}$$

This result demonstrates that block sparse vectors that are orthogonal remain nearly orthogonal after the application of $\mathbf{D}$.

**Lemma 2** [4]: Suppose that $\mathbf{D}$ has the Block RIP of order $K$ with isometry constant $\delta_d \in (0,1)$, $\Lambda \subseteq \{1,2,...,M\}$. If $|\Lambda| < K$, then $\forall \mathbf{u} \in \mathbb{R}^N$ with $\|\mathbf{u}\|_{2,0} \leq K - |\Lambda|$ and $\text{supp}_B(\mathbf{u}) \cap \Lambda = \varnothing$, we have that:

$$(1 - \frac{\delta_d}{1-\delta_d})\|\mathbf{u}\|_2^2 \leq \|\mathbf{A}_\Lambda \mathbf{u}\|_2^2 \leq (1+\delta_d)\|\mathbf{u}\|_2^2, \tag{10}$$

which means that, if $\mathbf{D}$ satisfies the Block RIP of order $K$ with isometry constant $\delta_d$, then $\mathbf{A}_\Lambda$ satisfies a restricted Block RIP of order $K - |\Lambda|$ with isometry constant $\frac{\delta_d}{1-\delta_d}$ (it is clear to see that $\frac{\delta_d}{1-\delta_d} < \delta_d$ for $\delta_d \in (0,1)$). Here the restriction is $\text{supp}_B(\mathbf{u}) \cap \Lambda = \varnothing$.

From Lemma 1 and Lemma 2, we prove Corollary 1 which is critical to our analysis below.

**Corollary 1:** Let $\mathbf{x} \in \mathbb{R}^N$ be given. If $\mathbf{D} \in \mathbb{R}^{L \times N}$ satisfies the Block RIP of order $\text{supp}_B(\mathbf{x}) + 1$ with isometry constant $\delta_d \in (0,1)$, where $N = Md$, then for all $j \in (1, M)$, we have

$$\|\mathbf{D}^T \mathbf{Dx}[j] - \mathbf{x}[j]\|_2 \leq \delta_d \|\mathbf{x}\|_2. \tag{11}$$

*Proof:* Let $\Gamma = \text{supp}_B(\mathbf{x}) \cup \{j\}$, so $|\Gamma| \leq \text{supp}_B(\mathbf{x}) + 1$. Let $\mathbf{Id}_\Gamma$ denotes the identity operator on $\mathbb{R}^\Gamma$, where $\mathbb{R}^\Gamma$ denotes that the space consisted of the columns whose block index belongs to $\Gamma$, the dimension of $\mathbb{R}^\Gamma$ is $|\Gamma|d$.

From the proof of Proposition 3.2 in [10], we know that $\|\mathbf{D}_\Gamma^T \mathbf{D}_\Gamma - \mathbf{Id}_\Gamma\|_{2 \to 2} = \sup_{\mathbf{y} \in \mathbb{R}^\Gamma, \|\mathbf{y}\|_2 = 1} \left| \|\mathbf{D}_\Gamma \mathbf{y}\|_2^2 - \|\mathbf{y}\|_2^2 \right|$, where $\|\mathbf{D}_\Gamma^T \mathbf{D}_\Gamma - \mathbf{Id}_\Gamma\|_{2 \to 2}$ is an operator and defined as $\|\mathbf{D}_\Gamma^T \mathbf{D}_\Gamma - \mathbf{Id}_\Gamma\|_{2 \to 2} (\mathbf{y}) \triangleq \|\mathbf{D}_\Gamma^T \mathbf{D}_\Gamma \mathbf{y} - \mathbf{y}\|_2$, $\mathbf{y} \in \mathbb{R}^\Gamma$. It is clear to see that

$$\|\mathbf{D}_\Gamma \mathbf{y}\|_2^2 = \|\mathbf{D}\bar{\mathbf{y}}\|_2^2 = \langle \mathbf{D}\bar{\mathbf{y}}, \mathbf{D}\bar{\mathbf{y}}\rangle, \|\mathbf{y}\|_2^2 = \|\bar{\mathbf{y}}\|_2^2 = \langle \bar{\mathbf{y}}, \bar{\mathbf{y}}\rangle = 1,$$

where $\bar{\mathbf{y}}$ is the continuation of $\mathbf{y}$ on $\mathbb{R}^N$, i.e.,

$$\bar{\mathbf{y}}[\Gamma(i)] = \mathbf{y}[i], \ i = 1, \cdots, |\Gamma|, \ \mathbf{y}[i] = \mathbf{0} \text{ if } i \notin \Gamma.$$

From Lemma 1, we have

$$\left\|\mathbf{D}_\Gamma^T \mathbf{D}_\Gamma - \mathbf{Id}_\Gamma\right\|_{2\to 2} = \sup_{\mathbf{y}\in\mathbb{R}^\Gamma, \|\mathbf{y}\|_2=1} \left|\langle \mathbf{D}\bar{\mathbf{y}}, \mathbf{D}\bar{\mathbf{y}}\rangle - \langle \bar{\mathbf{y}}, \bar{\mathbf{y}}\rangle\right| \le \delta_d \|\bar{\mathbf{y}}\|_2^2 = \delta_d.$$

Since $\mathrm{supp}_B(\mathbf{x}) \subseteq \Gamma$, we have

$$\left\|(\mathbf{D}^T\mathbf{D}\mathbf{x})|_\Gamma - \mathbf{x}|_\Gamma\right\|_2 = \left\|\mathbf{D}_\Gamma^T\mathbf{D}_\Gamma \mathbf{x}|_\Gamma - \mathbf{x}|_\Gamma\right\|_2 \le \delta_d \|\mathbf{x}|_\Gamma\|_2 = \delta_d \|\mathbf{x}\|_2.$$

Since $j\in\Gamma$, we have

$$\left\|\mathbf{D}^T\mathbf{D}\mathbf{x}[j] - \mathbf{x}[j]\right\|_2 \le \left\|(\mathbf{D}^T\mathbf{D}\mathbf{x})|_\Gamma - \mathbf{x}|_\Gamma\right\|_2 \le \delta_d \|\mathbf{x}\|_2. \quad\blacksquare$$

In what follows, we will prove Lemma 3 using Corollary 1, and from Lemma 3 the relationship between $\mathbf{h}^i$ and $\mathbf{x}$ can be determined.

**Lemma 3:** Suppose that $\Lambda \subseteq \{1,2,\ldots,M\}, \tilde{\mathbf{x}}\in R^N$ and $\mathrm{supp}_B(\tilde{\mathbf{x}}) \cap \Lambda = \varnothing$. Define

$$\mathbf{h} \triangleq \mathbf{A}_\Lambda^T \mathbf{A}_\Lambda \tilde{\mathbf{x}}. \tag{12}$$

If $\mathbf{D}$ satisfies the block RIP of order $\|\tilde{\mathbf{x}}\|_{2,0} + |\Lambda| + 1$ with isometry constant $\delta_d$, then $\forall j \notin \Lambda$ we have:

$$\|\mathbf{h}[j] - \tilde{\mathbf{x}}[j]\|_2 \le \frac{\delta_d}{1-\delta_d} \|\tilde{\mathbf{x}}\|_2. \tag{13}$$

*Proof:* From Lemma 2, we have that for all $\tilde{\mathbf{x}}$, $\mathbf{A}_\Lambda$ satisfies the restricted block RIP of order $\|\tilde{\mathbf{x}}\|_{2,0} + |\Lambda| + 1 - |\Lambda| = \|\tilde{\mathbf{x}}\|_{2,0} + 1$ with isometry constant $\frac{\delta_d}{1-\delta_d}$. Then from Corollary 1 and (12), we have (13) at once. $\blacksquare$

It is clear to see that $\mathbf{h}$ defined in (12) is almost the same as $\mathbf{h}^i$ in the $i$th iteration of Block OMP (see Algorithm 1 and (8)) Since the relationship between $\mathbf{h}^i$ and $\mathbf{x}$ is determined, we may derive a sufficient condition under which the identification step of Block OMP will succeed.

**Corollary 2:** Suppose that $\Lambda$, $\mathbf{D}$, $\tilde{\mathbf{x}}$ meet the assumptions specified in Lemma 3, and let $\mathbf{h}$ be as defined in (13). If

$$\|\tilde{\mathbf{x}}\|_{2,\infty} > \frac{2\delta_d}{1-\delta_d} \|\tilde{\mathbf{x}}\|_2, \tag{14}$$

where $\|\tilde{\mathbf{x}}\|_{2,\infty} \triangleq \max_j \|\tilde{\mathbf{x}}[j]\|_2$, then we have

$$\arg\max_j \|\mathbf{h}[j]\|_2 \in \mathrm{supp}_B(\tilde{\mathbf{x}}). \tag{15}$$

***Proof:*** If $j \notin \text{supp}_B(\tilde{\mathbf{x}})$, $\|\tilde{\mathbf{x}}[j]\|_2 = 0$. If $j \notin \text{supp}_B(\tilde{\mathbf{x}})$ and $j \notin \Lambda$, then from (13), we have that $\|\mathbf{h}[j]\|_2 \leq \frac{\delta_d}{1-\delta_d}\|\tilde{\mathbf{x}}\|_2$.

If $j \notin \text{supp}_B(\tilde{\mathbf{x}})$ but $j \in \Lambda$, then from (12) and the definition of $\mathbf{A}_\Lambda$, we have that $\|\mathbf{h}[j]\|_2 = 0$. So $\forall j \notin \text{supp}_B(\tilde{\mathbf{x}})$, $\|\mathbf{h}[j]\|_2 \leq \frac{\delta_d}{1-\delta_d}\|\tilde{\mathbf{x}}\|_2$. From (14), we have that $\exists j_0 \in \text{supp}_B(\tilde{\mathbf{x}}), s.t. \|\tilde{\mathbf{x}}[j_0]\|_2 > \frac{2\delta_d}{1-\delta_d}\|\tilde{\mathbf{x}}\|_2$. From (13) and the triangle inequality, we have that $\|\mathbf{h}[j_0]\|_2 > \frac{\delta_d}{1-\delta_d}\|\tilde{\mathbf{x}}\|_2$, so $\arg\max_j \|\mathbf{h}[j]\|_2 \in \text{supp}_B(\tilde{\mathbf{x}})$. ∎

By choosing $\delta_d$ small enough, it is possible to guarantee that the condition (15) is satisfied. Lemma 4 below may help us to find an appropriate $\delta_d$.

**Lemma 4:** $\forall \mathbf{x} \in R^N, \|\mathbf{x}\|_{2,\infty} \geq \frac{\|\mathbf{x}\|_2}{\sqrt{\|\mathbf{x}\|_{2,0}}}$, where $\|\mathbf{x}\|_{2,\infty} \triangleq \max_{1 \leq j \leq M} \|\mathbf{x}[j]\|_2^2$.

***Proof:*** $\|\mathbf{x}\|_{2,\infty}^2 = \max_{1 \leq j \leq M} \|\mathbf{x}[j]\|_2^2 = \frac{\|\mathbf{x}\|_{2,0} \max_{1 \leq j \leq M} \|\mathbf{x}[j]\|_2^2}{\|\mathbf{x}\|_{2,0}} \geq \frac{\sum_{j=1}^{\|\mathbf{x}\|_{2,0}} \|\mathbf{x}[j]\|_2^2}{\|\mathbf{x}\|_{2,0}} = \frac{\|\mathbf{x}\|_2^2}{\|\mathbf{x}\|_{2,0}}$, so $\|\mathbf{x}\|_{2,\infty} \geq \frac{\|\mathbf{x}\|_2}{\sqrt{\|\mathbf{x}\|_{2,0}}}$. ∎

Finally, using above results, we have the following main conclusion.

**Theorem 1:** Suppose that $\mathbf{D}$ satisfies the Block RIP of order $K+1$ with isometry constant $\delta_d < \frac{1}{2\sqrt{K}+1}$, then $\forall \mathbf{x} \in R^N$ and $\|\mathbf{x}\|_{2,0} \leq K$, Block OMP can exactly recover $\mathbf{x}$ from $\mathbf{y} = \mathbf{D}\mathbf{x}$ in $K$ steps.

***Proof:*** The theorem is proved by induction. At the first iteration, i.e., $l = 0$, $\mathbf{h}^0 = \mathbf{D}^T \mathbf{r}^0 = \mathbf{D}^T \mathbf{y} = \mathbf{D}^T \mathbf{D}\mathbf{x}$. From the definition of $\mathbf{A}_\Lambda$ we know that $\mathbf{D} = \mathbf{A}_\varnothing$. Since $\|\mathbf{x}\|_{2,0} \leq K$, from Lemma 4, we have that $\|\mathbf{x}\|_{2,\infty} \geq \frac{\|\mathbf{x}\|_2}{\sqrt{K}}$. Since $\delta_d < \frac{1}{2\sqrt{K}+1}$, we have $\frac{2\delta_d}{1-\delta_d} < \frac{1}{\sqrt{K}}$, and hence $\|\tilde{\mathbf{x}}\|_{2,\infty} > \frac{2\delta_d}{1-\delta_d}\|\tilde{\mathbf{x}}\|_2$. From Corollary 2, we have that $\arg\max_j \|\mathbf{h}^0[j]\|_2 \in \text{supp}_B(\mathbf{x})$.

Suppose that at the $i$th step, i.e., $l = i-1$, the conclusion in Theorem 1 is correct, i.e., $\Lambda^i \in \text{supp}_B(\mathbf{x})$ and $|\Lambda^i| = i$. Then when $l = i$, $\mathbf{h}^i = \mathbf{D}^T \mathbf{r}^i$, from the definition of $\mathbf{A}_\Lambda$ and (8), we have that $\mathbf{h}^i = \mathbf{A}_{\Lambda^i}^T \mathbf{A}_{\Lambda^i} \mathbf{x} = \mathbf{A}_{\Lambda^i}^T \mathbf{A}_{\Lambda^i} \tilde{\mathbf{x}}^i$, where $\tilde{\mathbf{x}}^i$ is defined as: $\tilde{\mathbf{x}}^i|_{\Lambda^i} = 0$, $\tilde{\mathbf{x}}^i|_{(\Lambda^i)^c} = \mathbf{x}|_{(\Lambda^i)^c}$. Since that $\text{supp}_B(\tilde{\mathbf{x}}^i) \cap \Lambda^i = \varnothing$, $\Lambda^i \subseteq \text{supp}_B(\mathbf{x})$, $\text{supp}_B(\tilde{\mathbf{x}}^i) \subseteq \text{supp}_B(\mathbf{x})$, $\|\mathbf{x}\|_{2,0} \leq K$ and $|\Lambda^i| = i$, we have that $\|\tilde{\mathbf{x}}^i\|_{2,0} \leq K - i$. Since $\mathbf{D}$ satisfies the Block-RIP of order $K+1$ with isometry constant $\delta_d < \frac{1}{2\sqrt{K}+1}$ and $K+1 = (K-i)+i+1 \geq \|\tilde{\mathbf{x}}^i\|_{2,0} + |\Lambda^i| + 1$, from

Lemma 4, we have

$$\|\tilde{\mathbf{x}}^i\|_{2,\infty} \geq \frac{\|\tilde{\mathbf{x}}^i\|_2}{\sqrt{\|\tilde{\mathbf{x}}^i\|_{2,0}}} \geq \frac{\|\tilde{\mathbf{x}}^i\|_2}{\sqrt{K-i}} \geq \frac{\|\tilde{\mathbf{x}}^i\|_2}{\sqrt{K}} > \frac{2\delta_d}{1-\delta_d}\|\tilde{\mathbf{x}}^i\|_2.$$

From Corollary 2, we have

$$j_0 = \arg\max_j \|\mathbf{h}^i[j]\|_2 \in \text{supp}_B(\tilde{\mathbf{x}}^i) \subseteq \text{supp}_B(\mathbf{x}),$$

and since $j_0 \in \text{supp}_B(\tilde{\mathbf{x}}^i)$ and $\text{supp}_B(\tilde{\mathbf{x}}^i) \cap \Lambda^i = \emptyset$, we have $j_0 \notin \Lambda^i$. $\Lambda^{i+1} = \Lambda^i \cup \{j_0\}$, so $|\Lambda^{i+1}| = i+1$.

When $l = \|\mathbf{x}\|_{2,0} - 1$, we have that $|\Lambda^l| = \|\mathbf{x}\|_{2,0}$ and $\mathbf{r}^l = \mathbf{A}_{\Lambda^l}\tilde{\mathbf{x}}^l$. From the definition of $\tilde{\mathbf{x}}^l$, we have $\tilde{\mathbf{x}}^l = 0$. So $\mathbf{r}^l = 0$, $\mathbf{y} = \mathbf{D}\mathbf{x}^l$.

Suppose that $\exists \mathbf{x}_1^l, \mathbf{x}_2^l$, $\text{supp}_B(\mathbf{x}_1^l) \subseteq \Lambda^l$, $\text{supp}_B(\mathbf{x}_2^l) \subseteq \Lambda^l$ s.t. $\mathbf{y} = \mathbf{D}\mathbf{x}_1^l = \mathbf{D}\mathbf{x}_2^l$, and we know that $\text{supp}_B(\mathbf{x}_1^l - \mathbf{x}_2^l) \subseteq \Lambda^l$, so $\|\mathbf{x}_1^l - \mathbf{x}_2^l\|_{2,0} \leq |\Lambda^l| \leq K$. From the Block RIP of $\mathbf{D}$, we know that $\|\mathbf{x}_1^l - \mathbf{x}_2^l\|_2^2 \leq \|\mathbf{D}(\mathbf{x}_1^l - \mathbf{x}_2^l)\|_2^2 = 0$, so $\mathbf{x}_1^l = \mathbf{x}_2^l$, which means the solution of Block OMP is unique. ∎

If $\mathbf{x}$ is treated as a conventional $Kd$ sparse vector without exploiting knowledge of the block sparse structure, a sufficient condition for exact recovery using OMP is that $\mathbf{D}$ satisfies the RIP of order $Kd+1$ with isometry constant $\delta < \frac{1}{3\sqrt{Kd}}$. It is clear to see that the condition for Block OMP is looser compared with that for OMP. Moreover, Block OMP need $K$ steps to recover $\mathbf{x}$ while OMP need $Kd$ steps, so Block OMP is faster thanks to the prior information of block sparsity.

When $d=1$ (i.e., the block size is equal to 1), the block sparse signal recovery problem becomes general sparse signal recovery problem. According to Theorem 1, we have that $\mathbf{D}$ is needed to satisfy the RIP of order $K+1$ with isometry constant $\delta < \frac{1}{2\sqrt{K}+1}$. This is different from the condition in [4] where $\mathbf{D}$ is needed to satisfy the RIP of order $K+1$ with isometry constant $\delta < \frac{1}{3\sqrt{K}}$. The reason is that in the proof of Theorem 3.1 in [4], $\delta < \frac{1}{3\sqrt{K}}$ is required to guarantee $\frac{2\delta}{1-\delta} < \frac{1}{\sqrt{K}}$, whereas $\delta < \frac{1}{2\sqrt{K}+1}$ is enough actually as a guarantee.

## IV. Conclusion

A greedy algorithm for block sparse signal, Block OMP algorithm is discussed in this paper. We analyze the recovery performance of Block OMP using Block RIP, and we prove Lemma 3, Lemma 4, Corollary 1 and Corollary 2 in Section III and at last prove Theorem 1 which states that if the dictionary matrix $\mathbf{D}$ satisfies the Block RIP of order

$K+1$ with isometry constant $\delta_d < \dfrac{1}{2\sqrt{K}+1}$, Block OMP can exactly recover block $K$-sparse signal in no more than $K$ steps, which is looser than the exact recovery condition using the conventional OMP. Moreover, the iteration steps using Block OMP are less than using OMP.